\documentclass[twocolumn]{webofc}
\usepackage[varg]{txfonts}   
\usepackage{booktabs}
\idline{}
\woctitle{FUSION23}
\wocname{EPJ Web of Conferences}
\makeatletter
\def\NAT@def@citea{\def\@citea{\NAT@separator}}
\makeatother 

\begin{document}
\title{Barrier distribution extraction via Gaussian process regression}
%
%

\author{\firstname{Kyle} \lastname{Godbey}\inst{2}\fnsep\thanks{\email{godbey@frib.msu.edu}}
}

\institute{Facility for Rare Isotope Beams, Michigan State University, East Lansing, Michigan 48824, USA
          }

\abstract{
This work presents a novel method for extracting potential barrier distributions from experimental fusion cross sections. We utilize a simple Gaussian process regression (GPR) framework to model the observed cross sections as a function of energy for three nuclear systems. The GPR approach offers a flexible way to represent the experimental data, accommodating potentially complex behavior without introducing strong prior assumptions. This method is applied directly to experimental data and is compared to the traditional direct extraction technique. We discuss the advantages of GPR-based barrier distribution extraction, including the capability to quantify uncertainties and robustness to noise in the experimental data.
}
\maketitle
\section{Introduction} 
\label{intro}
The probability that two nuclei will fuse in a reaction at a given energy is sensitive to both the initial character of the collision partners as well as the processes that take place during the reaction itself.
As such, it is natural to expect that fusion reaction studies should contain some measure of information regarding the intricate details of nucleus-nucleus interactions and the interplay of nuclear properties and the fusion process.
To study this interplay, however, extracting this information in a way that is less reliant on prior model assumptions is crucial for a deeper understanding of the dynamics of low-energy heavy-ion collisions.

While conventional techniques to study the nucleus-nucleus interaction often impose a predefined barrier model to interpret fusion reactions, a more flexible, data-driven approach may reveal insights not readily captured by parameterized potentials.
The simplest approach to study these features arising from the data alone is to look at derived quantities that only involve experimentally measured values, such as the fusion cross section.
Such techniques have been extremely successful in extracting insights from the raw experimental data for a range of nuclei covering the chart of nuclides.
The following work explores this approach and attempts to extend the method via a principled regression scheme while providing a Bayesian estimate of regression uncertainty.
Section~\ref{sec-1} will detail the formalism of the barrier distribution as well as the machine learning approach used to interpolate the data. Section~\ref{sec-2} presents results for the $^{16}$O+$^{208}$Pb, $^{40}$Ca+$^{40}$Ca, $^{40}$Ca+$^{48}$Ca, and $^{48}$Ca+$^{48}$Ca systems.
Finally, a brief summary is given in Sec.~\ref{sec-3} along with some perspective on future improvements to the method.

\section{Formalism}
\label{sec-1}
\subsection{Barrier Distributions}

In a theoretical context the extraction of barrier properties can be relatively straightforward.
In some approaches the fusion barrier itself is the primary object and its form and behavior is derived, primarily, empirically~\cite{akyuz1981,siwek-wilczynska2004}.
Even in theoretical frameworks that do not explicitly impose the existence of a heavy-ion interaction barrier, one can be extracted.
Regardless of if the potential is extracted in a purely static manner (e.g.~\cite{misicu2006,simenel2017}) or if microscopic dynamics are introduced in a time-dependent framework (e.g.~\cite{umar2006b,washiyama2008}), many methods have been developed to extract an effective one dimensional potential from a many-body description of fusion.
Despite the clear reduction in complexity, the extraction, analysis, and interpretation of these heavy-ion fusion barriers has yielded an incredible amount of insight into the intricate dynamics at play in the fusion process.
As an example from time-dependent density functional theory (TDDFT), the impact of transfer on sub-barrier fusion cross sections can be immediately seen with an expressive microscopic model and a means by which to derive an effective interaction potential~\cite{godbey2017}.
This is naturally a model dependent result, but it allows one to explore the correlation between separate phenomena and confront them explicitly to experimental data.

From the experimental point of view, no unique fusion barrier actually exists in nature.
That the fusion of two quantum many-body systems can be reduced to a two-body interaction potential and reproduce experimental data to great success is remarkable and highlights the dominating impact of the Coulomb interaction paired with the largely mean-field driven dynamics of fusion at low energies.
A method to extract the details of this potential directly from the data was proposed in Ref.~\cite{rowley1991} and it was immediately successfully in revealing deep insights into the physics of fusing systems~\cite{dasgupta1998}.
The so-called barrier distribution itself amounts to the second derivative of the energy-weighted total cross section,
\begin{equation}
    \label{eq:barrierdist}
    D(E) = \frac{d^2(E\sigma)}{dE^2}.
\end{equation}
In Eq.~\ref{eq:barrierdist} and the coming results we have neglected the $1/\pi R^2$ normalization that is typically present.
As the barrier distribution relies on the second derivative of the cross section, one requires high energy resolution data in order to keep the numerical error arising from the finite differences under control.
Next we introduce a way to mitigate this issue via a regression scheme to resample the excitation function to some arbitrary energy resolution, similar to the local regression in Ref.~\cite{Scamps2018a}.

\subsection{Gaussian Process Regression}

Gaussian Process Regression (GPR) provides a powerful and versatile Bayesian framework for regression analysis~\cite{rasmussen2005}.
GPR is a non-parametric technique that allows for the modeling of complex, non-linear dependencies within arbitrary datasets.
A Gaussian process (GP) defines a distribution over a collection of possible functions, where any finite subset follows a multivariate Gaussian distribution.
The core assumptions of a GPR approach is the selection and fitting of the kernel functions (also called covariance functions), which quantify the similarity between data points and determine how predictions are formed.
The choice of kernel function(s) in the GP model is quite important and should be chosen with the expected data variations in mind.
For more details and a comprehensive discussion of possible kernel functions see Ref.~\cite{duvenaud_2014}.

For the preliminary study in this work, a standard radial basis function (RBF) kernel is used with bounds placed on the minimum length scale determined from the energy resolution available in the experimental data.
In this work a single RBF kernel is used, limiting the expressivity of the resultant GPR statistical model.
A more comprehensive study of GP kernels for fusion excitation functions will be the focus of a future study.
Errors at the design points are directly from the experimental data, meaning the larger uncertainty in certain energy regions will naturally be encoded in the GPR model.
Another limitation of the current work is that there are assumed to be no errors in the domain of the function.
These so called measurement errors are small (or unreported) for the systems chosen here, though in the event that they are large one should consider a more sophisticated approach.
To extract the barrier distributions, derivatives are taken using standard finite difference on a fine mesh.
For all results presented here, hundreds samples are drawn of the fusion excitation functions and derivatives are taken individually.
By plotting all distributions derived from the sampled GPs together, a sense of the regression uncertainty can be quantified.

\section{Results}
\label{sec-2}

We now investigate the performance of the GPR model on real data from a variety of systems.
The systems chosen for this study include the heavily asymmetric $^{16}$O+$^{208}$Pb and well as the symmetric and nearly-symmetric $^{40}$Ca+$^{40}$Ca, $^{40}$Ca+$^{48}$Ca, and $^{48}$Ca+$^{48}$Ca.
The motivation behind these nuclei are to cover even-even, closed-shell systems that exhibit differing levels of bulk nucleonic transfer.
The calcium systems in particular were recently shown to be good constraints on model uncertainties in dynamical simulations given their similar structures besides the neutron skin of $^{48}$Ca~\cite{godbey2022}.
Furthermore, systems that have at least one high quality dataset with a reasonable energy resolution are important to validate the GPR model with the traditional pointwise derivative.

\begin{figure}[!htb]
\centering
  \includegraphics[width=0.95\columnwidth]{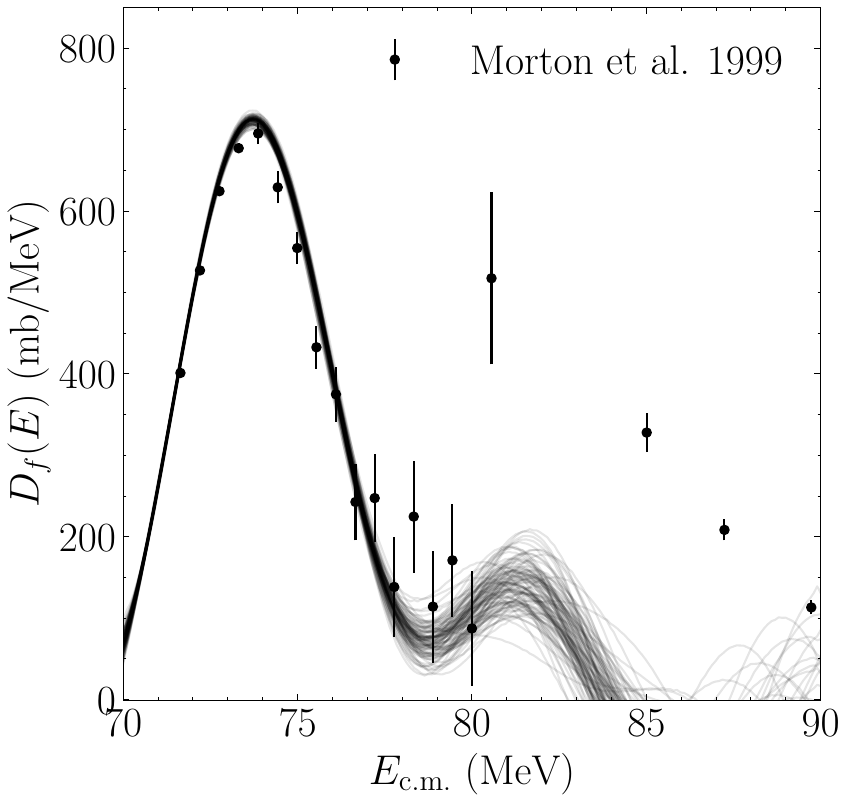}
  \caption{\protect Barrier distribution for the $^{16}$O$+^{208}$Pb system. Points are taken from the discrete derivative of the experimental data from Ref.~\cite{morton1999} and lines are samples from the GPR analysis of the same data.}
  \label{fig1}
\end{figure}

In Fig.~\ref{fig1} the $^{16}$O+$^{208}$Pb system is shown.
In this case, the original data from Ref.~\cite{morton1999} was sufficiently high resolution and had small enough errors to capture the broad barrier peak around 74~MeV quite well.
These results are also largely insensitive to the choice of $\Delta E$ in Eq.~\ref{eq:barrierdist}.
The GPR samples, shown as thin black lines, agree very well for the primary peak, exhibiting a small uncertainty until the data becomes more sparse and the errors grow around 77~MeV.
At large energies the pointwise derivative begins to deviate significantly due to the relatively sparse data in this region and it is very likely that the uncertainty on the point is severely underestimated.

\begin{figure}[!htb]
\centering
  \includegraphics[width=0.95\columnwidth]{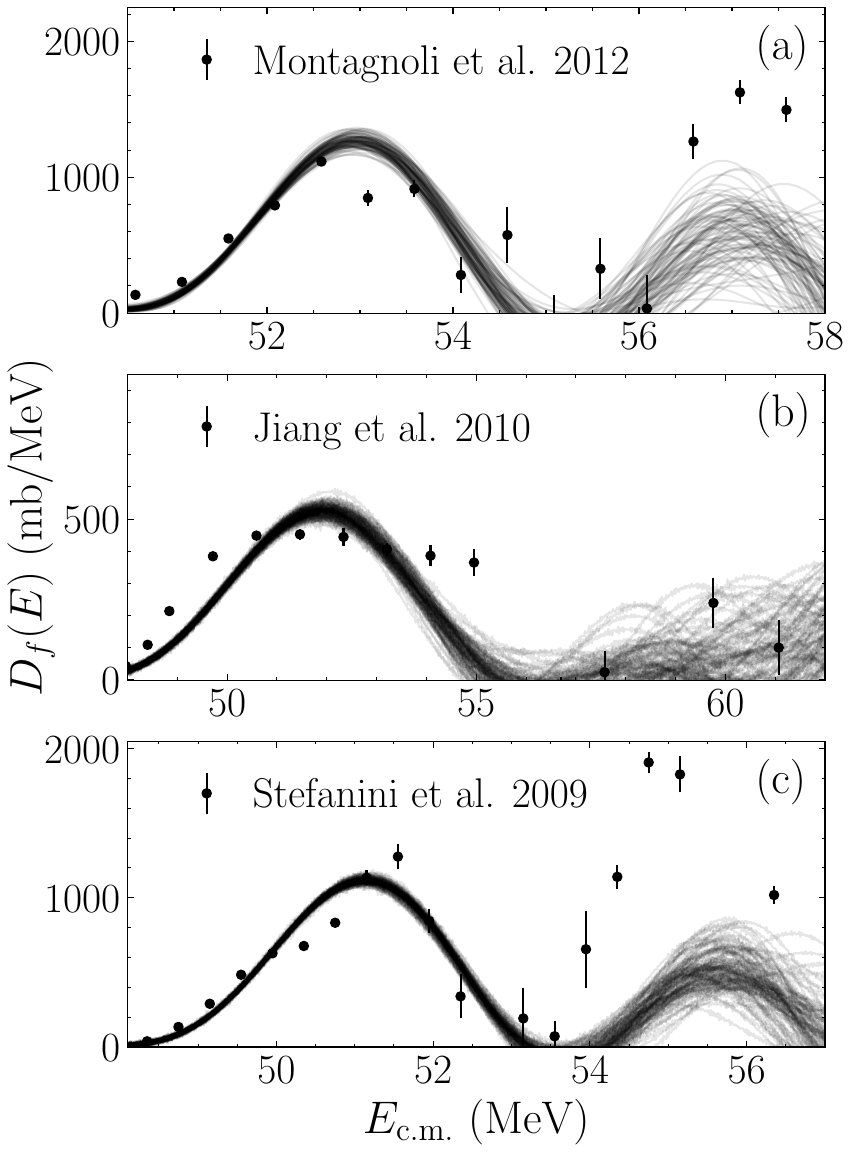}
  \caption{\protect Similar to Fig.~\ref{fig1}. Barrier distributions for the (a) $^{40}$Ca$+^{40}$Ca~\cite{montagnoli2012}, (b) $^{40}$Ca$+^{48}$Ca~\cite{jiang2010}, and (a) $^{48}$Ca$+^{48}$Ca~\cite{stefanini2009} systems.}
  \label{fig2}
\end{figure}

Figure~\ref{fig2} shows the results for the calcium systems.
For all systems the barrier distributions from the GPR model of the cross sections agrees quite well with the pointwise values near the primary barrier peak.
At higher energies the structure of the secondary peaks are captured relatively well, though the magnitude seems to be underestimated, particularly with the symmetric $^{48}$Ca system.
There are minor deviations in the structure of the primary peak that likely arise from the simple statistical model chosen to represent the cross sections as well as the fact that the GPR model is optimized to the cross sections in a linear scale.
This last explanation is likely the primary reason the GPR model is losing fidelity near the barrier energy as that is the point that the cross sections start to exponentially decay, meaning the model is not capturing any fine details in the region.
The issue of multiscale behavior in general is a strong motivation for exploring more expressive GP kernels, even if that amounts to simple additive and multiplicative combinations of squared-exponential kernels.

A common feature of all results is the fact that, at higher energies, the spread in the GPR-derived distributions grows substantially.
This is largely due to a combination of a growing uncertainty in the original GPR model of the cross section as one approaches the edge of available data and the intrinsic nature of the energy weighted derivative inducing a larger spread.
For studying fine features in the barrier distribution at larger energies, such as the oscillations in Ref.~\cite{Rowley15}, one should carefully consider the underlying statistical model in the GPR and the relative sparsity of the underlying data before applying the approach.

\section{Summary}
\label{sec-3}

In this work we have presented a data-driven procedure to represent fusion excitation functions with the only model assumption being the underlying statistical model generating the data.
In scenarios where experimental data is relatively sparse, irregularly spaced, or lacking measurement errors Gaussian processes provide a flexible statistical framework to learn the structures in the underlying data while providing a Bayesian estimation of regression uncertainties.
This flexibility is particularly attractive in reactions involving collisions partners that are far from stability as, despite the power of new facilities like the Facility for Rare Isotope Beams, the expected rates of reaccelerated nuclei near the dripline are such that low statistics should be expected for near and subbarrier fusion reactions.
Given that these nuclei often represent those that have the largest uncertainties in model predictions, a robust mechanism to deal with sparse data could act as a vital constraint for future Bayesian studies.

While the current work limits exploration to the simplest of the GP kernel functions, this is an important area of future investigation.
Either through handcrafted kernel selection or a more automated procedure~\cite{duvenaud_2014}, the multiscale nature of fusion cross sections warrant a more complicated statistical model that the ones explored here if the goal is to simultaneously explore energies far above and deep below the barrier.
For studies focused solely on near and above barrier phenomena, the choice to use a single, expressive kernel with the combination of an optional, additional noise term is well-motivated.
The squared-exponential kernel itself is attractive for derived quantities like the barrier distributions, given their infinitely differentiable nature.
This smoothness has been beneficial for the heavy systems considered here, though recent studies on light-ion fusion has shown their potential to inform theoretical approaches to understanding the dynamics of fusion~\cite{desouza2023search}.
The presence of higher frequency behavior arising from the $\ell$-barriers and knock-out channels could motivate a more `jagged' structure, such as the behavior in the Mat\'ern kernel.
Motivated in part by recent light ion measurements~\cite{Asher21a,Hudan23}, another area that warrants investigation is the proper way to handle measurement errors in the domain as energies are rarely known to absolute precision, particularly in experiments based on a multi-sampling ionization chamber.
Such issues are open questions in the GP literature~\cite{zhou2023}, motivating interdisciplinary collaboration to ensure that nuclear physics continues to exploit the cutting edge of technology.

\section*{Acknowledgements}

I thank the organizers of the FUSION23 conference for fostering a vibrant, collegial atmosphere where some of the concepts and conclusions in this work were fleshed out.
This work has been supported by the U.S. Department of Energy under award number DE-NA0004074 (NNSA, the Stewardship Science Academic Alliances program).

%
\bibliography{VU_bibtex_master,ExtraBib}

\end{document}